\documentclass{PoS}

\usepackage{graphicx}

\usepackage{overpic}

\usepackage{multicol}

\title{The VERITAS Dark Matter Program}

\ShortTitle{VERITAS Dark Matter Program}

\author{\speaker{Benjamin Zitzer}\\
        McGill University\\
        E-mail: \email{bzitzer@physics.mcgill.ca}}

\author{for the VERITAS Collaboration\\
        Affiliation\\
        E-mail: \email{bzitzer@physics.mcgill.ca}}

\abstract{In the cosmological paradigm, cold dark matter (DM) dominates the mass content of the Universe and is present at every scale. Candidates for DM include many extensions of the Standard Model with weakly interacting massive particles (WIMPs) in the mass range from ~10 GeV to greater than 10 TeV. The self-annihilation or decay of WIMPs in astrophysical regions of high DM density can produce secondary particles including very-high-energy (VHE; E>100 GeV) gamma rays. VERITAS, an array of atmospheric Cherenkov telescopes, sensitive to VHE gamma rays in the 85 GeV-30 TeV energy range, has been utilized for indirect DM searches. Astrophysical candidates for indirect DM detection by VERITAS are dwarf spheroidal galaxies (dSphs) of the Local Group and the Galactic Center, among others. This is a report on the current status of VERITAS observations of dark matter targets and the results of a combined search of four dSphs.}

\FullConference{35th International Cosmic Ray Conference – ICRC217-\\
		10-20 July, 2017\\
		Bexco, Busan, Korea}

\begin{document}

\section{Introduction}

The search for standard model particles resulting from the annihilation of dark matter provides an important complement to the efforts of direct searches for dark matter interactions and searches for dark matter production at particle accelerators. Among the theoretical candidates for the dark matter particle above a few GeV, Weakly Interacting Massive Particles (WIMPs) are well motivated \cite{1996PhR...267..195J,2003NuPhB.650..391S} as they naturally provide the measured present day cold dark matter density \cite{2014A&A...571A..16P,1979ARNPS..29..313S,1965PhL....17..164Z,zel1965advances,1966PhRvL..17..712C}.  In such models, the WIMPs either decay or annihilate into standard model particles that produce mono-energetic gamma-ray lines and/or a continuum of gamma rays with energies up to the dark matter particle mass. 

Good targets for searches with gamma-ray instruments are those with a large inferred dark matter (DM) density and low astrophysical backgrounds; i.e. those that would not produce gamma rays through conventional processes. Dwarf Spheroidal Galaxies (dSphs) fit these criteria since they are close by (tens to hundreds of kpc away) and with no theoretical basis for emitting very-high-energy (VHE) gamma rays. The center of our galaxy is also a tantalizing target since it is even closer, about $\sim$8 kpc away. However the center of the galaxy has many astrophysical backgrounds including diffuse emission along the galactic plane \cite{2006Natur.439..695A}; the radio source at the center of our galaxy, SgrA*, is also a very bright VHE emitter. The metric for a good DM target is the $J$-factor (or $D$ factor in the case of decay) defined as:

\begin{equation}
J = \int\int\rho(l,\Omega)^{2}dld\Omega \\
\quad\quad D = \int\int\rho(l,\Omega)dld\Omega,
\end{equation}
where $\rho$ is the DM density and $l$ is the direction along the line-of-sight. This work will focus on DM annihilations. We will also refer to the $J$-profile which is equal to $J$-factor per solid angle. The $J$-factor is directly proportional to the expected gamma-ray flux from DM annihilations,

\begin{equation}
 \frac{d\Phi}{dE} = \frac{\langle\sigma\nu\rangle}{8\pi M^{2}}J(\Delta\Omega)\sum_{i}B_{i}\frac{dN_{\gamma,i}}{dE},
\end{equation}
where $M$ is the WIMP particle mass, $dN_{\gamma}/dE$ is the energy spectrum of a single annihilation with branching ratio $B_{i}$, and $\langle\sigma\nu\rangle$ is the velocity-averaged annihilation cross section.

VERITAS (Very Energetic Radiation Imaging Telescope Array System) is an array of four imaging atmospheric Cherenkov telescopes (IACTs), each 12 m in diameter, located at the Fred Lawrence Whipple Observatory in southern Arizona, USA.  VERITAS is sensitive to gamma rays from approximately 85 GeV (after camera upgrade) to greater than 30 TeV with a typical energy resolution of $15-25\%$  and an angular resolution (68\% containment) of $<$0.1 degrees per event. The flux sensitivity of the standard analysis is such that a source with a flux of order of 1\% of the Crab Nebula flux can be detected in approximately 25 hours of observation.

\section{Combined Analysis of Dwarf Galaxies from 2007 to 2013}

From the beginning of four-telescope operations in 2007 to the summer of 2013, five dSphs in the northern hemisphere have been observed: Segue~1, Ursa Minor, Draco, Bo\"{o}tes and Willman~1. No significant gamma-ray excess was observed in the direction of any of the five dSphs. A dark matter search was performed with 216 hours of VERITAS data, including 92 hours on the dSph Segue~1. Table 1 summarizes the analysis results of each of the five dSphs. 

The 95\% confidence level limits on the thermally-averaged annihilation cross section are shown in Figure 1. The search for DM annihilation and the limits on the cross section were derived from the Event Weighting method presented in Geringer Sameth et. al. \cite{2015PhRvD..91h3535G}. Willman~1 was excluded from the combined search and limits since the $J$-factor can not be calculated reliably due to irregular kinematics \cite{2015ApJ...801...74G}. Additionally, Bonnivard et. al. \cite{2015MNRAS.446.3002B} have pointed out the possibility of contamination of the stellar samples used to perform the Jeans analysis which could bias $J$-profiles towards larger annihilation signals, possibly by orders of magnitude. Therefore, two different limits are shown in each panel in Figure 1: including Segue~1 and without. Each panel in Figure~1 represents a 100\% branching ratio into a given standard model particle. The limits are presented as a band with the representing the $\pm$1$\sigma$ systematic uncertainty of the $J$-profile. 

\begin{table*}[t]
	\begin{center}
	\scalebox{0.8}{
	\begin{tabular}{ l | c | c | c | c | c | c | c | c}
	DSph & $N_{ON}$ & $N_{OFF}$ & $\bar{\alpha}$ & Significance & $N^{95\%}$ & $\Phi^{95\%}$                          & Distance   & $\log_{10} J(0.17^{\circ})$ \\
	          & [counts]   & [counts]     &                        & [$\sigma$]  & [counts]        & [10$^{-12}$cm$^{2}$s$^{-1}$] & [kpc] & [GeV$^{2}$ cm$^{-5}$] \\
	\hline
	Segue 1   	 & 	15895  &	120826	&	0.131	&  0.7   	& 235.8	& 0.34  & 23  & 19.2$^{+0.3}_{-0.3}$ \\
	Draco      	 &   	 4297   &	 39472	&  	0.111	& -1.0	& 33.5	& 0.15  & 76  & 18.3$^{+0.1}_{-0.1}$ \\
	Ursa Minor &      4181   &	 35790	&	0.119	& -0.1   	& 91.6	& 0.37  & 76  & 18.9$^{+0.3}_{-0.3}$ \\
	Bo{\"o}tes 1 &  	 1206   &	 10836	&	0.116	& -1.0 	& 34.5	& 0.40  & 66  & 18.3$^{+0.3}_{-0.4}$ \\
	Willman 1   &     1926   &	 18187	&	0.108	& -0.6    	& 23.5	& 0.39  & 38  & N/A\\
	\end{tabular}
	}
	\caption{DSph detection significance (generalized Li \& Ma method) \cite{1983ApJ...272..317L} and integral flux upper limit with 95\% confidence level above 300 GeV, assuming a spectral index of -2.4 using the Rolke method \cite{2005NIMPA.551..493R}. The last two columns are the heliocentric distance to each object and the inferred value of $J$-profile integrated within a cone with half-angle of $0.17^{\circ}$ (i.e. over the ON region); errors denote the 16th and 84th percentiles on the posterior \cite{2015ApJ...801...74G}. } 
	\label{tab:dwarfsresult}
	\end{center}
\end{table*}

\begin{figure*}
\centering
	\includegraphics[scale=0.55]{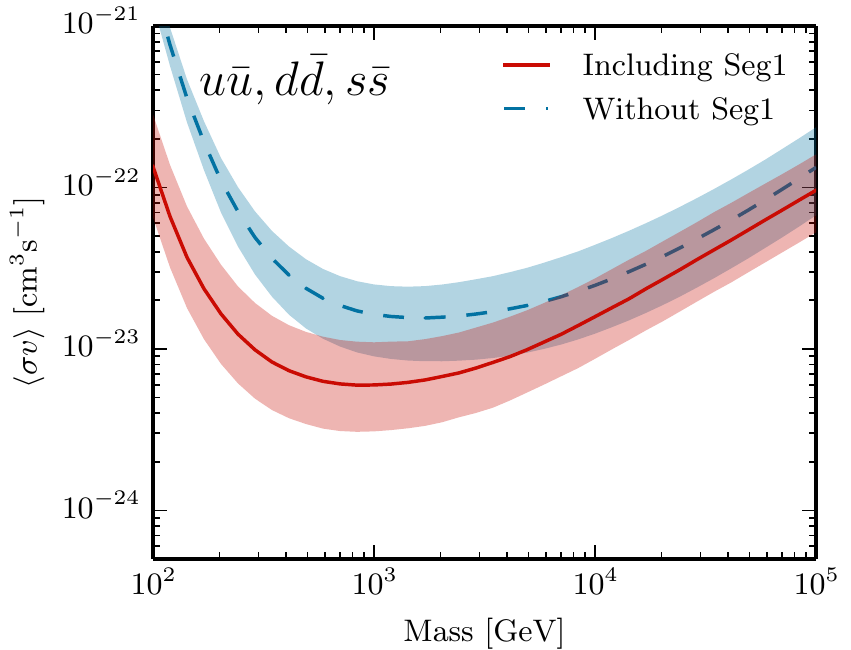}
	\includegraphics[scale=0.55]{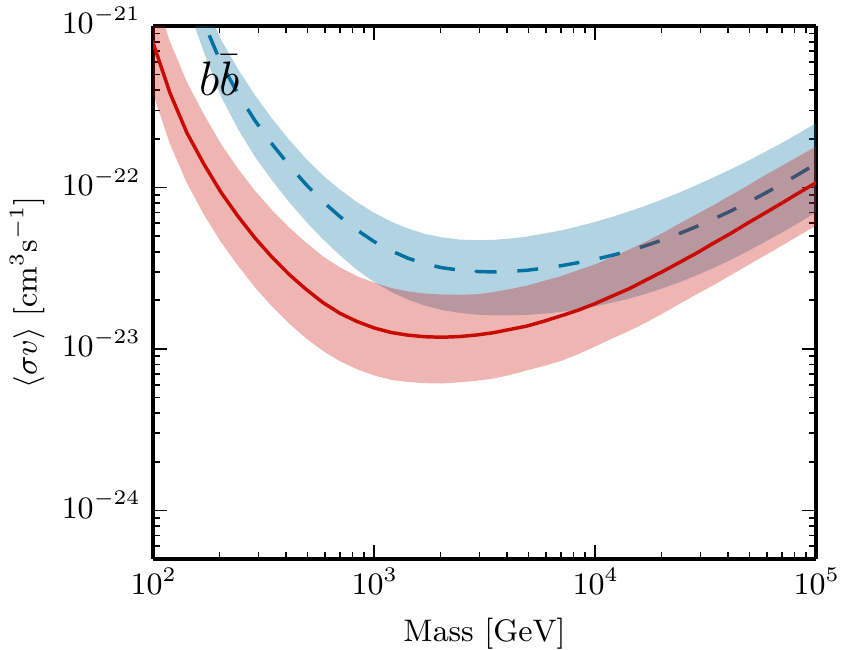}
	\includegraphics[scale=0.55]{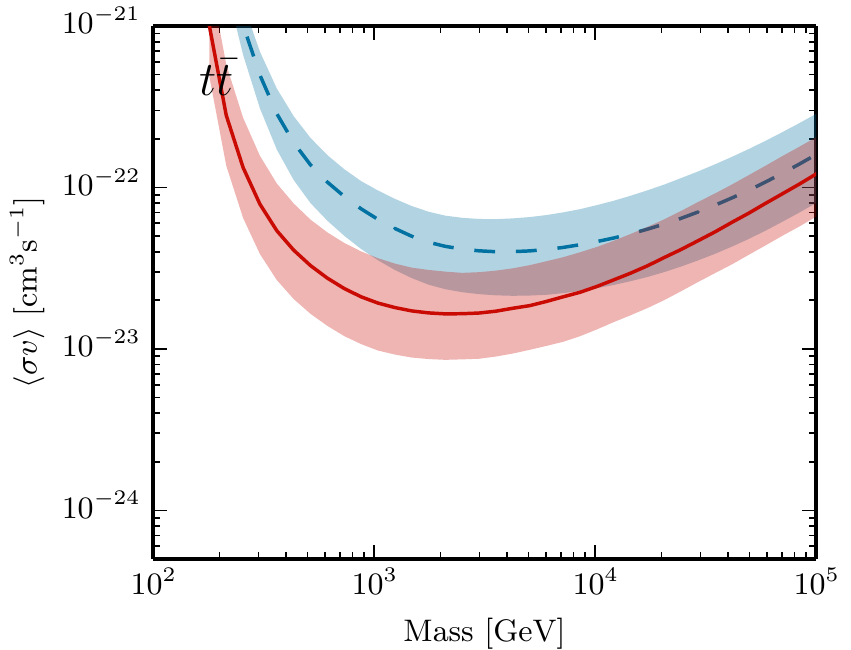}\\
	\includegraphics[scale=0.55]{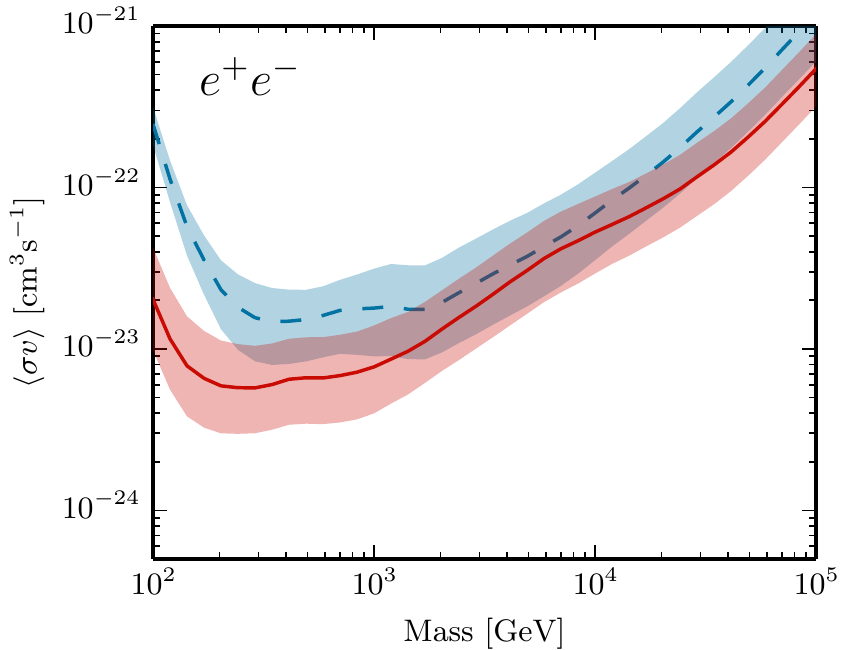}
	\includegraphics[scale=0.55]{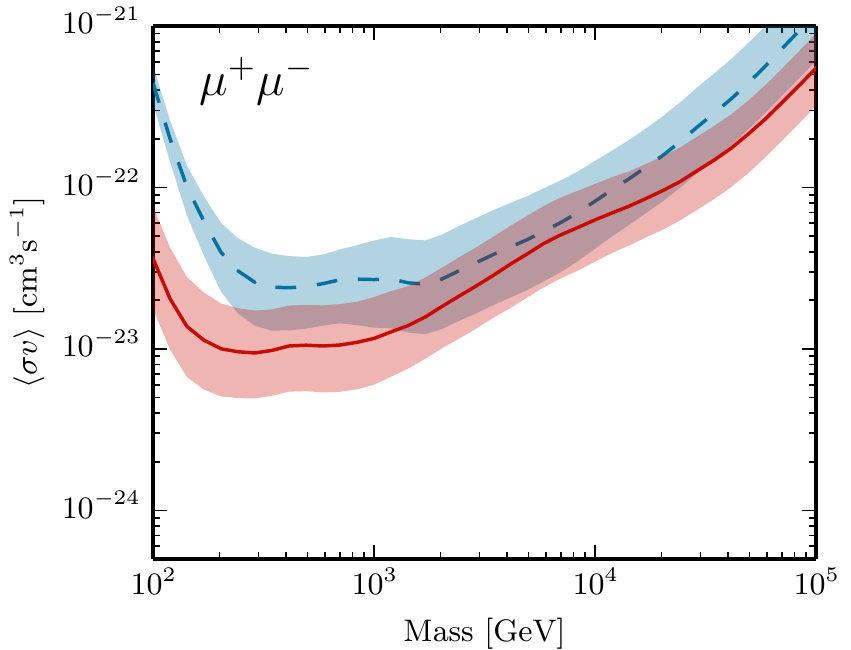}
	\includegraphics[scale=0.55]{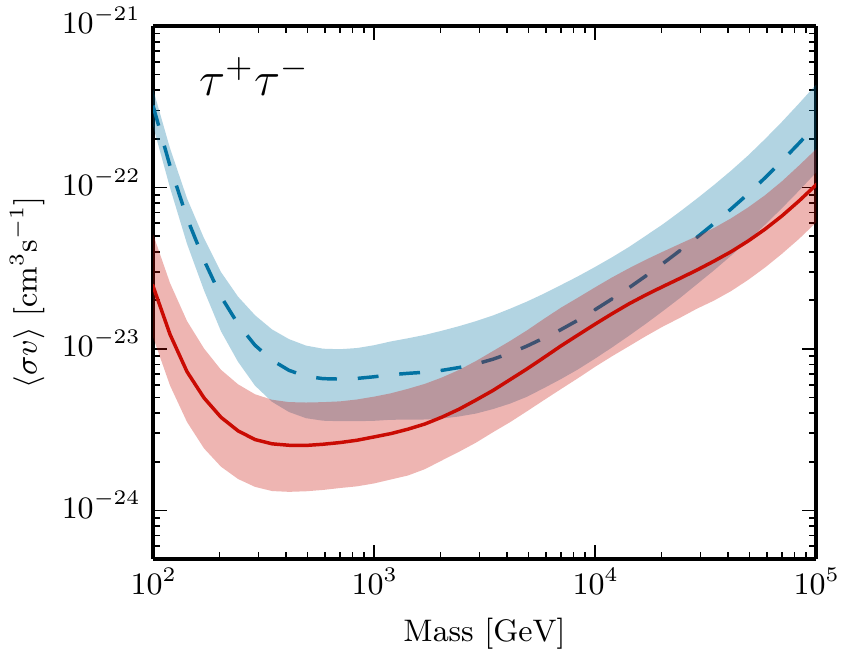} \\
	\includegraphics[scale=0.55]{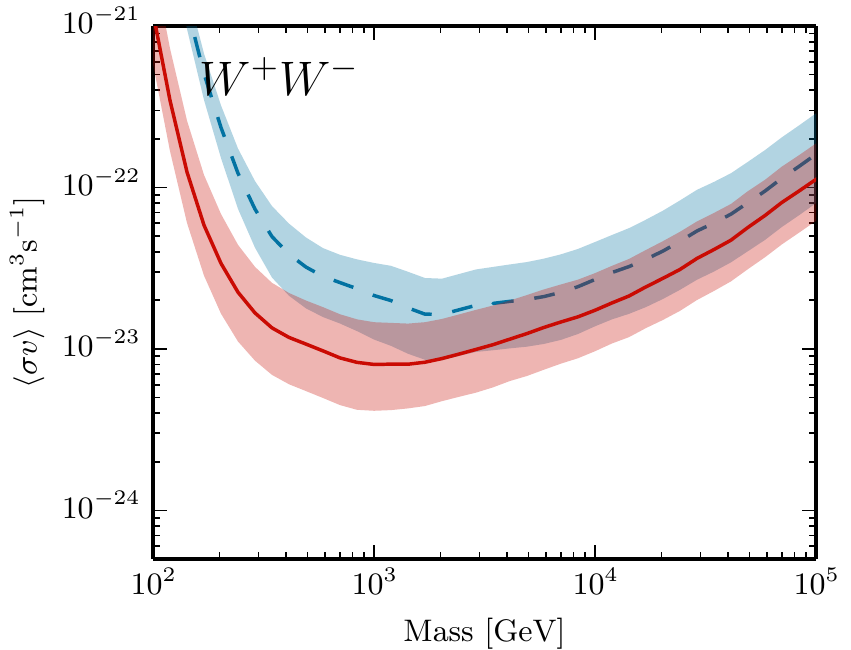}
	\includegraphics[scale=0.55]{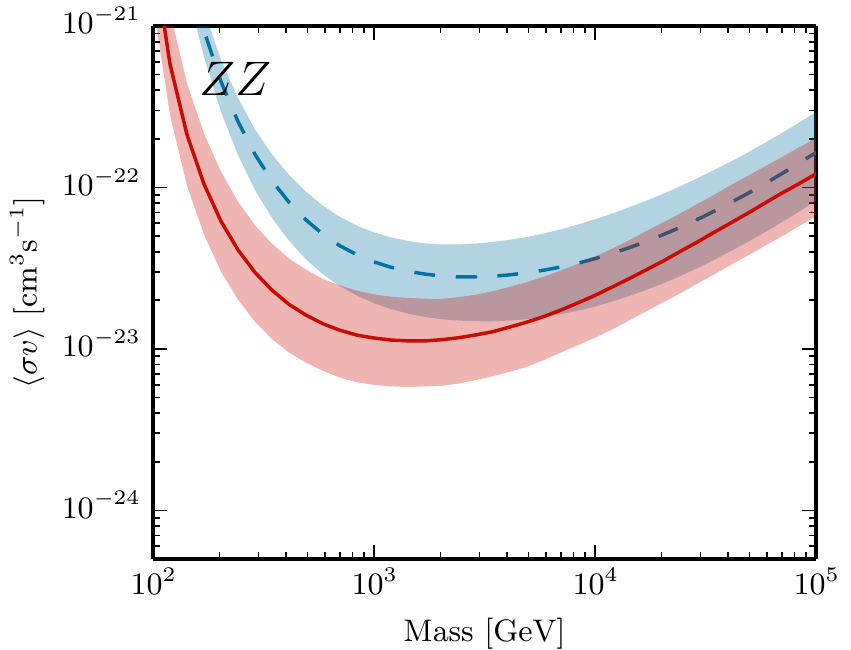}
	\includegraphics[scale=0.55]{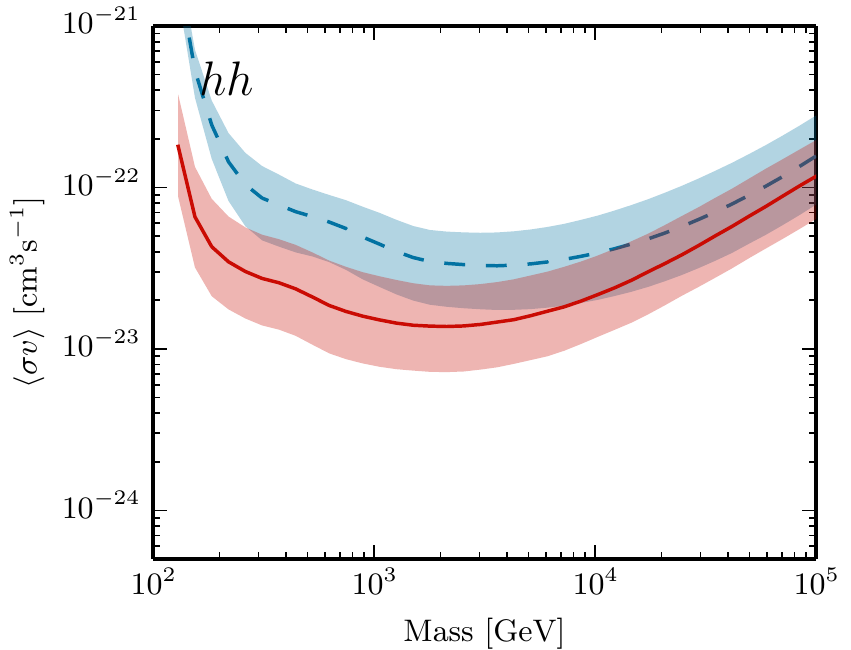} 
	\includegraphics[scale=0.55]{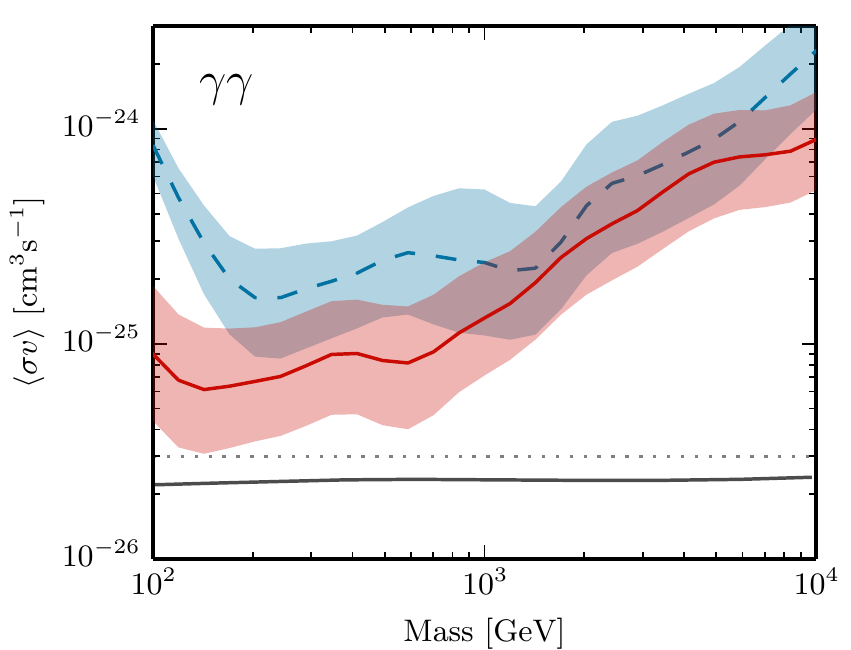}
\caption{Annihilation cross section limits from the joint analysis of dSphs.  The shaded bands are the systematic $\pm1\sigma$ uncertainty in the limit derived from many realizations of halo $J$-profiles of the dSphs consistent with kinematic data. The solid line depicts the median of this distribution of limits over the halo realizations with all dSphs except Willman~1. The dashed line depicts the median limits of the distribution of limits without Segue~1 and Willman~1. }
\label{fig:combinedlimits_syst}
\end{figure*}

\section{A Decade of Dwarf Galaxy Observations: 2007 to 2017}

The indirect detection of dark matter is one of the collaboration's key science projects, and dwarf galaxies are an integral part of the program. Because VERITAS is not a survey instrument, we have to prioritize the DM targets within our observing budget. DSphs observed by VERITAS are subdivided into two categories: `Deep Exposure' dSphs and `Survey' dSphs. `Deep Exposure' dSphs are those with the best $J$-factors in the literature which get the majority of the observing time. This time is typically split between the ultra-faint and classical dSphs. This is done as a way to ensure that the overall DM program is not severely penalized in case one particular dSph falls out of favor as a good indirect DM target. Most of the dSphs that do not have deep exposures are in the survey for much shorter exposures, typically a few hours per year. The survey provides DM sensitivity to nearly all dSphs in the northern hemisphere as a form of insurance in case misunderstanding of velocity dispersion data leads to a dSph having a much larger $J$-factor than otherwise indicated. The survey and the deep exposure dSphs can be also be combined or `stacked' into a more sensitive limit than any one individual dSph as shown in the previous section. 

Table 2 lists all dSph observations since the beginning of four-telescope operations in 2007 to mid-2017. The exposure accumulated on for each dSph are divided into observing epochs: the original array configuration (v4), after the move of Telescope 1 (v5) and after the camera upgrade (v6). See \cite{NaheeICRC2015} for details on the performance of the different epochs. The observation times in Table 2 do not include time intervals with bad weather or hardware malfunctions, but rather the total observing time so these numbers should be considered an upper limit to the possible exposure of each dSph. The analysis of many of these dSph data sets is still preliminary. 

\subsection{Ursa Major II}
 As shown in Table 2, a large amount of the total dSph observing time after the camera upgrade has been devoted to Ursa Major II.  Ursa Major II was discovered in SDSS data with follow-up observations from Subaru \cite{2006ApJ...650L..41Z}. It is one of the faintest known dSphs, with velocity dispersion measurements made with only 20 stars \cite{2012AJ....144....4M}, but with a large mass-to-luminosity ratio. $J$-factors for Ursa Major II typically indicate that it is one of the highest, often similar to or even exceeding Segue~1 (e.g. \cite{2015ApJ...801...74G,2015MNRAS.446.3002B}. 

VERITAS observed Ursa Major II between late 2013 and spring 2017.  After dead-time corrections and removal of time intervals with bad weather or hardware performance, a live time of 145 hours remain. Observation runs with fewer than all four telescopes were also removed from this analysis. The analysis of the data follows the same procedure as in \cite{VerDsph2017}. The \textit{hFit} algorithm is used for event reconstruction \cite{2012AIPC.1505..709C} and the crescent background method is used to subtract the remaining cosmic-ray background \cite{VerDsph2017}. An excess of 1$\sigma$ above the background was found at the Ursa Major II position. We report a flux upper limit of 1.4$\times$10$^{-8}$m$^{-2}$s$^{-1}$ above 200 GeV at a 95\% confidence level using the method of Rolke \cite{2005NIMPA.551..493R} assuming a powerlaw with index of -2.4.
 
Figure 2 shows preliminary annihilation cross section limits for Ursa Major II. A version of the full maximum likelihood introduced by Aleksi{\'c} et. al. \cite{2014JCAP...02..008A} was used for the search for DM annihilations and for determining the limit to the velocity-averaged cross section.  The $J$-profile is convolved with the VERITAS PSF, generated from monte carlo simulations has described in \cite{VerDsph2017}. We do this because the DM profile is wider than the PSF, particularly at lower energies. The median and $\pm$1$\sigma$ $J$-profile was provided in the supplemental material of Geringer-Sameth et al. \cite{2015ApJ...801...74G}. The annihilation spectrum is the one used from PPPC 4 DM ID \cite{2011JCAP...03..051C}. As shown in Figure 2, the VERITAS limit for Ursa Major II is more constraining on the cross section than the 216 hrs combined dSph \cite{VerDsph2017} limit at every DM mass by a factor of $\sim$30\% to more than a factor of 2, depending on DM mass. 

\begin{figure}
	\centering
	\begin{overpic}[scale=0.37]{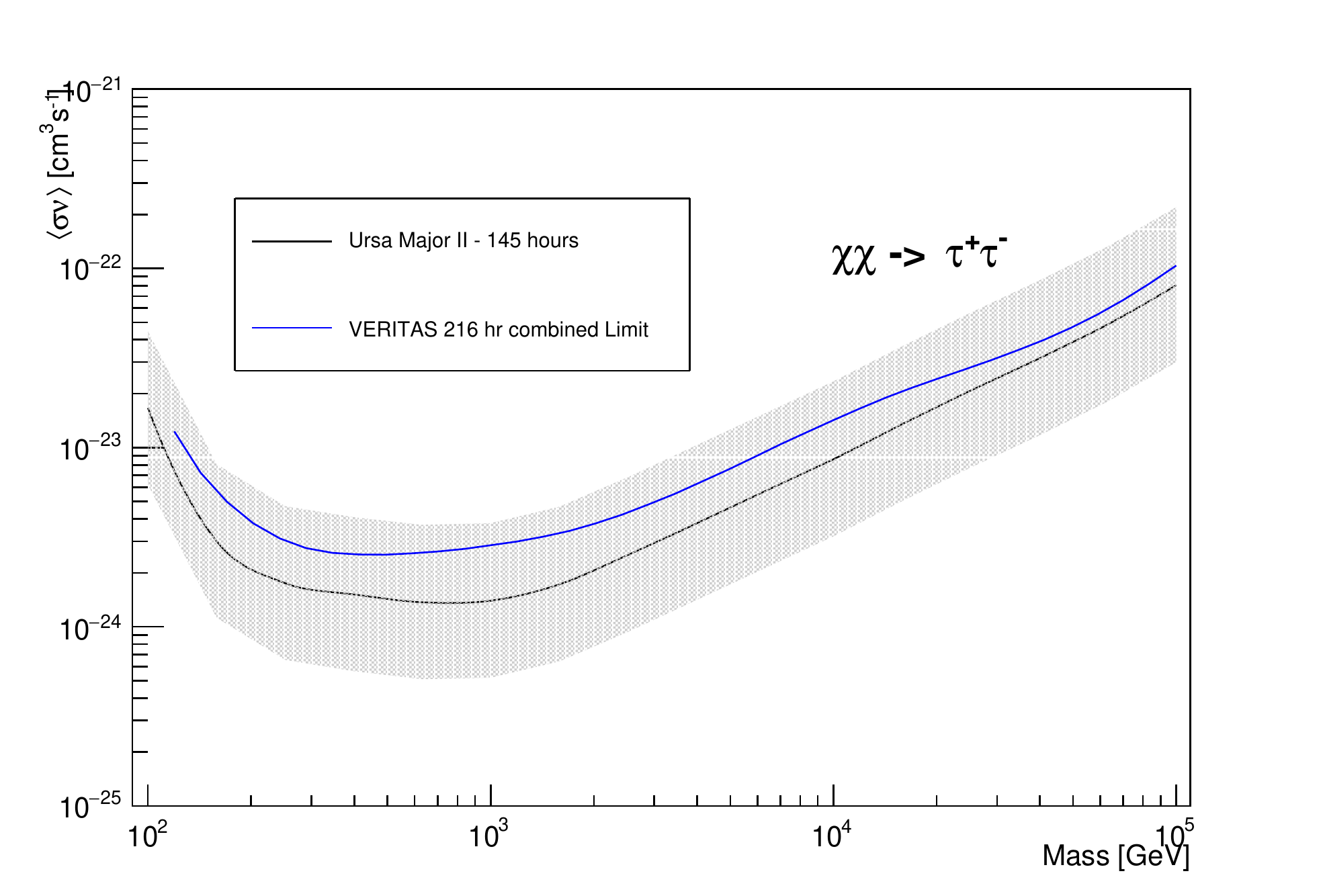}
	\put(20,12){\small\color{red} VERITAS - ICRC 2017}
	\end{overpic}
	\begin{overpic}[scale=0.37]{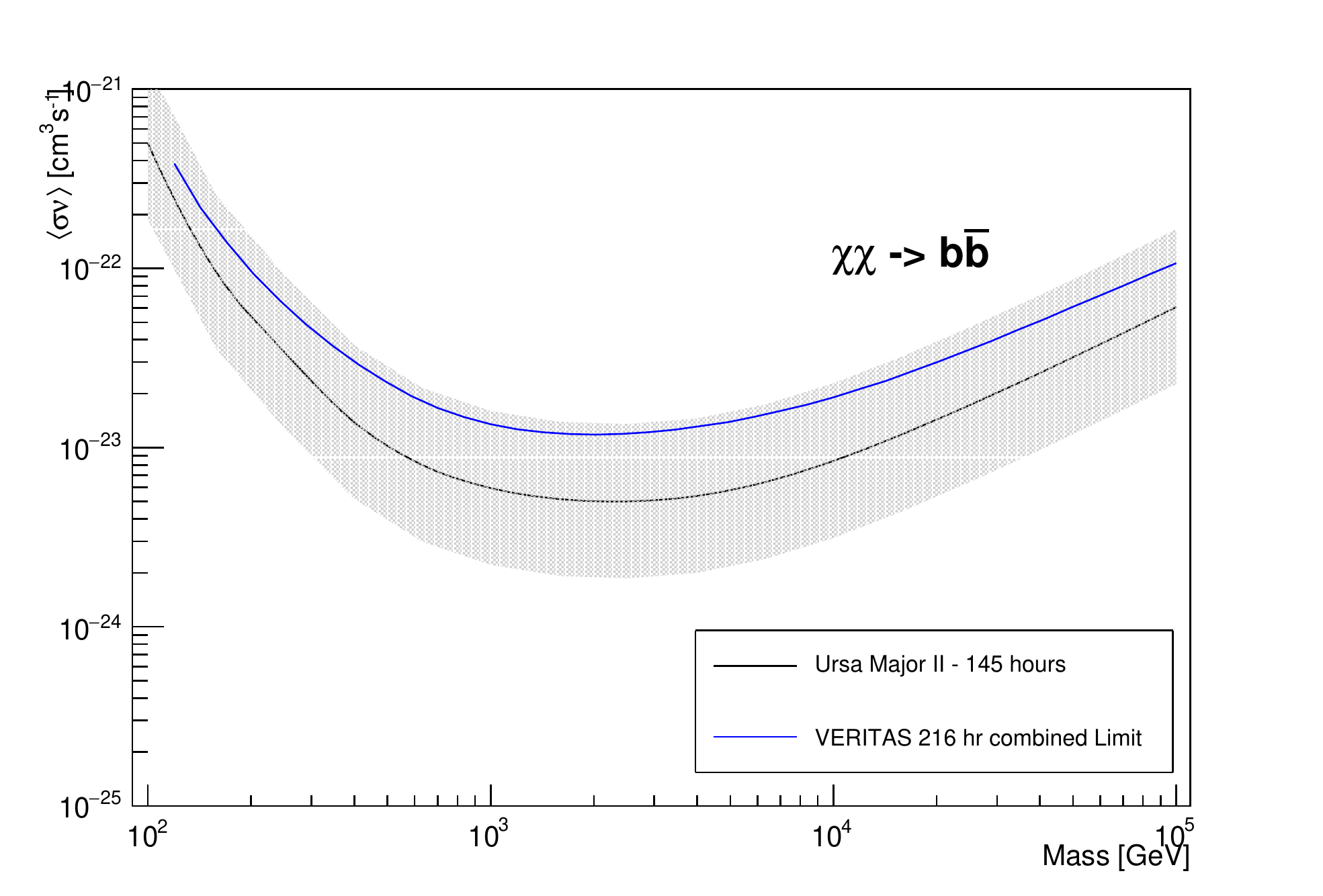}
	\put(20,52){\small\color{red} VERITAS - ICRC 2017}
	\end{overpic}

	\caption{Preliminary DM annihilation limits for 145 hours of Ursa Major II taken with VERITAS, assuming a 100\% branching ratio of annihilation into $\tau$ leptons (left) and $b$ quarks (right). The shaded band represents $\pm$1$\sigma$ error bands on the $J$-profile. } 
\end{figure}

\begin{table*}[t]
	\begin{center}
	\scalebox{0.75}{
	\begin{tabular}{ l | c | c | c | c | c | c | c }
	Dwarf	& $\log_{10}J_{1}(0.5^{\circ})$	& $\log_{10}J_{2}(0.5^{\circ})$	& $\log_{10}D_{1}(0.5^{\circ})$	& Exposure v4	& Exposure v5	& Exposure v6 	& Total Exposure \\
			& [GeV$^{2}$ cm$^{-5}$] 		& [GeV$^{2}$ cm$^{-5}$] 		& [GeV cm$^{-2}$]	& [min]	& [min]	& [min]	& [min] \\ 
	\hline
	Segue 1		& 19.4$^{+0.3}_{-0.4}$	&	17.0$^{+2.1}_{-2.2}$		&	18.0$^{+0.2}_{-0.3}$	&	0	&	6121	&	4921	&	11042	\\
	Ursa Major II	& 19.4$^{+0.4}_{-0.4}$	&	19.9$^{+0.7}_{-0.5}$	&	18.4$^{+0.3}_{-0.3}$	&	0	&	0	&	10869&	10869	\\	
	Ursa Minor 	& 18.9$^{+0.3}_{-0.2}$	&	19.0$^{+0.1}_{-0.1}$	&	18.0$^{+0.2}_{-0.1}$	&	711	&	2209	&	6844	&	9724		\\
	Draco		& 18.8$^{+0.1}_{-0.1}$	&	19.1$^{+0.4}_{-0.2}$	&	18.5$^{+0.1}_{-0.1}$	&	1169	&	2170	&	3435	&	6813		\\
	Coma Berencies	& 19.0$^{+0.4}_{-0.4}$	&	19.6$^{+0.8}_{-0.7}$	&	18.0$^{+0.2}_{-0.3}$	&	0	&	0	&	2204	&	2204		\\	
	Segue II		& 16.2$^{+1.1}_{-1.0}$	&	18.9$^{+1.1}_{-1.1}$	&	15.9$^{+0.4}_{-0.4}$	&	0	&	0	&	1128	&	1128		\\
	Bo{\"o}tes 1	& 18.2$^{+0.4}_{-0.4}$	&	18.5$^{+0.6}_{-0.4}$	&	17.9$^{+0.2}_{-0.3}$		&	960	&	0	&	0	&	960		\\
	Leo II 		& 18.0$^{+0.2}_{-0.2}$	& 	17.8$^{+0.2}_{-0.2}$		&	17.2$^{+0.4}_{-0.5}$		&	0	&	0	&	946	&	946		\\	
	Willman 1		& N/A 				&	N/A 					&	N/A					&	931	&	0	&	0	&	931		\\
	Triangulum II	& N/A 				&	N/A 					&	N/A 					&	0	&	0	&	909	&	909		\\
	Canes Ver. II 	& 17.7$^{+0.5}_{-0.4}$	&	18.5$^{+1.2}_{-0.9}$	&	17.0$^{+0.2}_{-0.2}$		&	0	&	0	&	864	&	864		\\
	Canes Ver. I 	& 17.4$^{+0.4}_{-0.3}$	&	17.5$^{+0.4}_{-0.2}$		&	17.6$^{+0.4}_{-0.7}$		&	0	&	0	&	850	&	850		\\	
	Hercules I 		& 16.9$^{+0.7}_{-0.7}$	&	17.5$^{+0.7}_{-0.7}$		&	16.7$^{+0.4}_{-0.4}$	&	0	&	0	&	794	&	794		\\
	Sextans I 		& 18.0$^{+0.2}_{-0.2}$	&	17.6$^{+0.2}_{-0.2}$		&	17.9$^{+0.1}_{-0.2}$		&	0	&	0	&	783	&	783		\\		
	Draco II		& N/A				&	N/A 					&	N/A					&	0	&	0	&	598	&	598		\\
	Ursa Major I 	& 17.9$^{+0.6}_{-0.3}$	&	18.7$^{+0.6}_{-0.5}$	&	17.6$^{+0.2}_{-0.4}$		&	0	&	0	&	482	&	482		\\
	Leo I 		& 17.8$^{+0.2}_{-0.2}$	& 	17.8$^{+0.5}_{-0.2}$		&	17.9$^{+0.2}_{-0.2}$		&	0	&	0	&	409	&	409		\\
	Leo V 		& 16.4$^{+0.9}_{-0.9}$	&	16.1$^{+1.2}_{-1.0}$	&	15.9$^{+0.5}_{-0.5}$	&	0	&	0	&	167	&	167		\\
	Leo IV 		& 16.3$^{+1.1}_{-1.7}$	&	16.2$^{+1.5}_{-1.6}$	&	16.1$^{+0.7}_{-1.1}$	&	0	&	0	&	151	&	151		\\

	\end{tabular}
	}
	\caption{Summary of Dwarf Spheroidal Galaxies observed by VERITAS between mid-2007 and mid-2017. For DM annihilation, two different $J$-factors are given over a integral solid angle of 0.5$^{\circ}$: those from Geringer-Sameth et al. 2015 \cite{2015ApJ...801...74G} ($J_{1}$) and those from Bonnivard et al. 2015 \cite{2015MNRAS.446.3002B} ($J_{2}$) . $D$-factors are also given over a integral solid angle of 0.5$^{\circ}$ from Geringer-Sameth et al. 2015  ($D_{1}$). Exposure times are given for the three major epochs of VERITAS operations: Before the move of telescope 1 (v4), after the move of telescope 1 but before the camera upgrade (v5) and after the camera upgrade (v6). See Park et al. 2015 \cite{NaheeICRC2015} for performance details and sensitivity curves.  }
	\end{center}
\end{table*}

\section{Galactic Center}


The center of our galaxy harbors a 4$\times$10$^{6}$ solar mass black hole, coincident with the radio source SgrA*. The region around the center of our galaxy is very complex in gamma rays. This region includes bright sources such as the supernova remnant G0.6+0.1 \cite{2016ApJ...821..129A}, diffuse emission along the plane as observed by HESS \cite{2006Natur.439..695A} and Fermi-LAT as well as the extended Fermi bubbles \cite{2010ApJ...724.1044S}. The existence of several possible astrophysical backgrounds makes finding a DM signal from WIMP annihilation a complicated but not impossible prospect. Because VERITAS is a Northern Hemisphere instrument, the Galactic Center never transits above 30$^{\circ}$ elevation. Observing at these elevations has the effect of raising the energy threshold and widening the PSF, but increasing sensitivity for higher energy showers. The degradation of the PSF is mostly offset by the displacement method of shower reconstruction \cite{2016ApJ...821..129A}. VERITAS is therefore the most sensitive instrument for observing the Galactic Center at energies greater than 2 TeV, with a PSF of 0.12$^{\circ}$ (68\% containment). 


\section{Dark Matter Subhalos}
Cosmological N-body simulations \cite{springel} indicate that DM halos contain significant substructures \cite{diemand}. Because of tidal disruption near the Galactic disk, most of the sub-halos
are thought to survive at high Galactic latitude. The lack of material in these regions prevents the DM over-densities from attracting enough baryonic matter to trigger star formation. DM clumps would therefore
be invisible to most astronomical observations from radio to X-rays. DM structures residing in the the Milky Way halo could potentially be nearby the Sun and therefore have a bright gamma-ray annihilation signal. These clumps would therefore only be visible at gamma-ray energies and not been in any other wavelength. 

The Second Fermi-LAT Catalog (2FGL) contains 576 high-energy gamma-ray sources detected by the LAT after the first 24 months of observations without any clear associations in any other wavelength. To identify the unassociated Fermi objects (UFOs) most likely to be DM clump candidates we adapt the selection criteria from \cite{nieto}, requiring that they are located outside of the Galactic Plane, that they be non-variable, they exhibit power law spectra and that they have no known counterparts. The two best candidates were selected for VERITAS observations: 2FGL J0545.6+6018 and 2FGL J1115.0-0701. It should be noted that at the time these objects were proposed for observation, that the 2FGL catalog was the most up-to-date Fermi-LAT catalog available. 2FGL J1115.0-0701 was observed for 13.8 hours and 2FGL J0545.6+6018 was observed for 8.5 hours. No emission was detected by VERITAS, giving flux upper limits 0.5\% and 0.6\% of the Crab nebula flux above 250 GeV, respectively. Later analysis of 2FGL J1115.0-0701 in Fermi-LAT showed that the flux was variable, excluding the possibility of it being a DM subhalo. Updated Fermi-LAT analysis of 2FGL J0546.6+6018 shows that a curved power law is actually favored over a simple power law, but DM annihilation to $b$ quarks or $W$ bosons also provide a good spectral fit. Optical follow-up observations were performed within the 95\% error region of 2FGL J0546.6+6018 by the 1.3m McGraw-Hill telescope located at Kitt Peak, AZ, USA. A previously undiscovered bright UV source was observed within the error circle, the analysis and interpretation of this is still ongoing. Swift-XRT observations show no new associations \cite{2013ApJS..207...28S}. 

\section{Summary and Future Work}
VERITAS has an active Dark Matter and Astroparticle group that leads this key science project of indirect DM detection for the collaboration. Continued observations of dSphs and the Galactic Center region are a crucial part of the VERITAS $\textit{long term plan}$ \cite{2015ICRC...34..868S}. Unfortunately, no potential subhalo candidates have been found in the 3FGL catalog. The Galactic Center region could potentially be the best indirect $\gamma$-ray source for dark matter discovery or stringent limits on annihilation cross sections at the highest masses. Future work for dwarf galaxies involve a new combined analysis with not only new data, but with other $\gamma$-ray instruments, namely Fermi-LAT and HAWC, which is still in the preliminary stages. New analysis techniques such as boosted decision trees (BDTs) \cite{2017APh....89....1K} and template reconstruction \cite{JodiTheseProceedings} have potential for improving the overall dark matter sensitivity of the experiment. Additionally many dSphs are extended for IACTs, so an extended source analysis could give a boost to DM sensitivity by as much as a factor of two in some cases. 

\section{Acknowledgments}
This research is supported by grants from the U.S. Department of Energy Office of Science, the U.S. National Science Foundation and the Smithsonian Institution, and by NSERC in Canada. We acknowledge the excellent work of the technical support staff at the Fred Lawrence Whipple Observatory and at the collaborating institutions in the construction and operation of the instrument. The VERITAS Collaboration is grateful to Trevor Weekes for his seminal contributions and leadership in the field of VHE gamma-ray astrophysics, which made this study possible.

\begin{multicols}{2}

\end{multicols}

\end{document}